\documentclass[onecolumn,preprintnumbers,amsmath,amssymb]{revtex4}
\usepackage{graphicx}
\usepackage{epsfig}
\usepackage{amsmath}
\usepackage{amsfonts}
\usepackage{amssymb}
\usepackage{textcomp}
\usepackage{dcolumn}
\usepackage{feynmf}

\begin{document}

\title{Quantum Theory of Ur Objects and General Relativity}

\author{Martin Kober}
\email{kober@th.physik.uni-frankfurt.de}

\affiliation{Institut f\"ur Theoretische Physik, Johann Wolfgang Goethe-Universit\"at, Max-von-Laue-Str.~1,
60438 Frankfurt am Main, Germany}

\begin{abstract}

The quantum theory of ur objects postulates that all existing physical objects and their properties are constructed from fundamental objects called ur objects being described by an element of a two dimensional complex Hilbert space. This approach is based on the assumption that quantum theory represents a theory being constitutive for human knowledge. Physical objects are characterized by the information one can gain from them being contained in the quantum state they are described by. Since every Hilbert space can be represented as a tensor product of two dimensional Hilbert spaces, one is led to the ur objects. According to this approach relativistic quantum fields and thus the existence of a Minkowski space-time are the consequence of an iteration of a quantization of binary alternatives. In the original formulation there was only obtained a description of quantum fields on a flat Minkowski space-time. In this work there is made the attempt to incorporate general relativity
and to describe the gravitational field within this approach. Thus the existence of a (3+1)-dimensional space-time in the sense of general relativity is assumed to be a consequence of quantum theory interpreted in an abstract sense.

\end{abstract}

\maketitle

\section{Introduction}

The reconciliation of quantum theory with general relativity is one of the most important and perhaps even the decisive problem in contemporary theoretical physics. There arises the question in which conceptual framework the gravitational field should be quantized. In this context it is one central task to understand the relation between the concepts of general relativity on the one hand and those of quantum theory on the other hand. This implies the question if one of these theories has to be assumed to be more fundamental or if the concepts of both theories are not general enough and one has to exceed the conceptual framework of both theories to obtain a real understanding of their relation to each other. Roughly spoken one could assert that quantum theory in its manifestation within relativistic quantum field theories the standard model of particle physics is based on provides the conceptual framework for the description of matter and its interactions on space-time and general relativity provides the conceptual framework for the description of the structure of space-time and gravity interpreted as a property of space-time. This duality between the geometry of space-time on the one hand and the existence of matter fields defined on space-time on the other hand occupied already Einstein when he was searching for a unified field theory. According to Einstein's attitude geometry was more fundamental than matter because in his opinion a geometrical description was nearer to pure mathematics. Therefore he tried to derive the properties of matter from a geometric description in accordance with the setting of gravity which led him to the attempt of a description of particles as singularities in space-time. However, if matter is described in the framework of quantum theory, or in the framework of relativistic quantum field theories to be more specific, similar to the geometric description of gravity it seems to be characterized in a very abstract and mathematical way. In the framework of relativistic quantum field theories particles are characterized as irreducible representations of the Poincare group \cite{Wigner:1939},\cite{Weinberg:1995mt}. Further, they are specified through their internal symmetries. Thus a particle is described and characterized by an abstract state within a Hilbert space and its corresponding symmetries. This means that in the context of a quantum theoretical description matter seems to be reduced more and more to the description of an abstract structure and this could be a hint that the argument of the beauty of a pure mathematical description of nature does not mandatorily lead to the attitude that the geometry of space-time has to be assumed to be more fundamental than the existence of matter fields. Concerning this central question it is further very important that quantum theory in its general Hilbert space formulation of Dirac and von Neumann \cite{Dirac:1958,vonNeumann:1932} without referring to particles or fields on space-time does not presuppose a position space. This could be a very important argument for the attitude that quantum theory is more fundamental than general relativity because its basic concepts are more abstract.\\

Carl Friedrich von Weizsaecker tried to reconstruct the physics of relativistic quantum field theories from quantum theory interpreted in a very principle sense \cite{Weizsaecker:1971},\cite{Weizsaecker:1985},\cite{Weizsaecker:1992},\cite{Weizsaecker:1955},\cite{Weizsaecker:1958a},\cite{Weizsaecker:1958b},\cite{Lyre:1994eg},\cite{Lyre:1995gm},\cite{Lyre:2003tr},\cite{Lyre:2004A}.
He assumed quantum theory to be constitutive for human perception and human knowledge. According to this approach physical objects are characterized through the information one can gain from them which is contained in their quantum states within an abstract Hilbert space. This means that the properties of physical objects are reduced to the concept of information which thus obtains the status of a substance, if one wants to express it in a philosophical language.
Since every Hilbert space can be represented as a tensor product of two dimensional Hilbert spaces, he was led to the quantum theory of fundamental objects described by a state within such a two dimensional space which he called ur objects (the denotation ur object is derived from the German prefix ur- which means something like original, elementary or primordial). Thus every object can be thought to be composed from fundamental objects being mathematically described by an element of a two dimensional complex Hilbert space which is nothing else than a Weyl spinor. The ur objects can be seen as indivisible objects in a rigorous but abstract and thus non spatial sense.
It would extend the scope of this paper to deal with C.F. von Weizsaecker's philosophical argumentations justifying quantum theory to be constitutive for human experience. It will be presupposed the basic assumption that abstract quantum theory without further assumptions yields the most fundamental description of nature accessible so far and it will be explored its consequences concerning an incorporation of general relativity.\\
With respect to this it is very important that the existence of a position space combined with the time direction to a (3+1)-dimensional space-time is assumed to be a consequence of this representation of quantum theoretical objects, since the symmetry properties of a two dimensional complex vector space correspond to the symmetry properties of a three dimensional real vector space. The last assertion has its origin in the local equivalence between the $SU(2)$, the group of unitary transformations in two dimensions with a determinant equal to one and the $SO(3)$, the symmetry group of rotations within a three dimensional space.
The above question concerning the relation between the description of space-time according to general relativity and the description of matter according to relativistic quantum field theories here appears in a very interesting way.\\
However, in the original setting of C.F. von Weizsaecker's approach, gravity was not incorporated. This means that the existence of a gravitational field as an additional structure of space-time is not derived so far. In this paper there is made the attempt to derive also the space-time description of general relativity from this approach and thus to incorporate gravity by using the translation gauge description of gravity and relating it to a description by spinor fields. The whole approach seems to be very plausible because of the reason that both theories one seeks to unify, quantum theory and general relativity, seem to imply that the property of the existence of a (3+1)-dimensional manifold representing space-time is the consequence of a representation of relationships between dynamical entities and has no reality for itself without a reference to them. In general relativity it is the property of general covariance or diffeomorphism invariance being related to the definition of inertia and absolute acceleration by the gravitational field as a dynamical entity and thus the lack of absolute objects referring to the space-time structure which suggests that space-time is no physical entity by itself \cite{Rovelli:1999hz},\cite{Rovelli:2006yt},\cite{Rovelli:2004},\cite{Lyre:2004B}.
With respect to quantum theory it is the mentioned fact that its abstract postulates as a theory of states and operators within a Hilbert space do not presuppose a position or a momentum space. These spaces just yield certain kinds of representations and this is related to the non locality of quantum theory.\\
The paper is structured as follows: At the beginning it is given an introduction to the quantum theory of ur objects and a description of the derivation of quantum theoretical field equations from an iteration of a quantization of binary alternatives in the sense of C.F. von Weizsaecker. The aim of the paper is to make the attempt to incorporate general relativity to this approach. This is tried in a way corresponding to the translation gauge description of gravity. Within this approach the gravitational field as tetrad field appears as gauge potential which has to be expressed by ur objects. At the end cosmological questions are considered.

\section{Reconstruction of Physics and the Quantum Theory of Ur Objects}

\subsection{General Ideas and Philosophical Preliminary}

According to C.F. von Weizsaecker's attempt of a reconstruction of physics from constitutive postulates about human knowledge, quantum theory is interpreted in a very principle sense \cite{Weizsaecker:1971},\cite{Weizsaecker:1985},\cite{Weizsaecker:1992},\cite{Weizsaecker:1955},\cite{Weizsaecker:1958a},\cite{Weizsaecker:1958b},\cite{Lyre:1994eg},\cite{Lyre:1995gm},\cite{Lyre:2003tr},\cite{Lyre:2004A}. Within this interpretation a physical object is characterized through the information one can gain from it. This leads to a description by n-fold alternatives of possible results of measurements as a general scheme. According to the postulated structure of time which presupposes that the result of a possible experiment in the future is not determined one is led to a non classical logic.
Already John von Neumann assumed in his book "The mathematical foundations of quantum mechanics" that quantum theory needs a new logic, called quantum logic, where a statement can also have other values than true or false \cite{vonNeumann:1932}. This means that the logical law called "tertium non datur" is not assumed to be valid any more. The Hilbert space structure of quantum theory is assumed to describe such a logical structure of n-fold alternatives which is assumed to be the structure of statements referring to the structure of time according to C.F. von Weizsaecker's philosophy of physics. In such a description one is led to probabilities for the elements of the n-fold alternative and one obtains the Hilbert space structure of what C.F. von Weizsaecker calls abstract quantum theory. This means that the fact that a physical object is characterized by its state within an abstract Hilbert space in quantum theory is interpreted in such a way that this state in Hilbert space represents the information which can be obtained from it. The state determines the probabilities which replace the statements true and false in a classical theory. It is logically trivial that every alternative can be represented as a combination of many binary alternatives. According to this the m-dimensional Hilbert space $\mathcal{H}$ of an arbitrary object can be described as the tensor product of two dimensional complex Hilbert spaces (which shall be denoted by $\mathbb{C}^2$)

\begin{equation}
\mathcal{H}^m \subseteq T^n=\bigotimes_n \mathbb{C}^2,\quad m<2^n.
\end{equation}
These two dimensional objects represented by states within the corresponding Hilbert spaces are assumed to be the fundamental entities of nature and are called ur objects by C.F. von Weizsaecker. They are the simplest objects even thinkable in any quantum theoretical description of nature.
It is now very important that, as already mentioned, the postulates of quantum theory in the general Hilbert space formulation do not presuppose a position space or a momentum space although they do presuppose the existence of time.
According to the quantum theory of ur objects the existence of a three dimensional position space attached with time to a Minkowski space is a derived quantity. It is a consequence of the properties of the two dimensional complex Hilbert space of ur objects. Such a Hilbert space has the symmetry group $SU(2)$, the unitary group in two dimensions with a determinant equal to one being isomorphic to the $SO(3)$ being the rotation group within three dimensions. 
An element of a two dimensional complex Hilbert space, a two dimensional Weyl spinor, can be mapped directly to a Minkowski vector by using the Pauli matrices the unit matrix in two dimensions included, whereas the components obtained from the Pauli matrices correspond to the spatial components having $SO(3)$ symmetry and the component obtained from the unit matrix corresponds to the time component.
The correspondence between Minkowski vectors and Weyl spinors is reflected by the fact that the group of general linear transformations with a determinant equal to one $SL(2,\mathbb{C})$ corresponds to the homogeneous Lorentz group $SO(3,1)$.
If all ur objects a physical system consists of are transformed with the same element of $SU(2)$ or $SL(2,\mathbb{C})$ respectively, the physics is not changed.  This is assumed to be the origin of the existence of a Minkowski space-time as a (3+1)-dimensional manifold with its corresponding symmetries all physical objects are located in which appears therefore as a consequence of quantum theory. Thus the nature of space-time and its three dimensional spatial part have their origin according to C.F. von Weizsaecker in a representation of the abstract Hilbert spaces of quantum theory as a tensor product of binary alternatives having this symmetry.
It is now possible or even necessary to iterate the description of quantum theory. The alternatives which can be constructed from binary alternatives and reflecting a quantum logic have to be treated quantum theoretically by themselves.
In this sense also the quantum state corresponding to an ur object has to be treated as an element of an ensemble of possible states building a new alternative on a higher logical level. This leads to the concept of multiple quantization, the iteration of a quantum theoretical description.\\
According to such an iteration of quantization applied to binary alternatives it is possible to obtain quantum theoretical field equations. One begins with a simple binary alternative. Quantization of such an alternative leads to a spinor which can be mapped to a Minkowski vector. Another quantization, corresponding to the usual first quantization, leads to a wave function depending on the Minkowski vector and thus corresponds to relativistic particle quantum mechanics. A further iteration of quantization leads to the second quantization of quantum field theory. Thus the field quantization is interpreted as an iteration of the quantization of point particles leading to wave functions. This is in accordance with the well known property that multi particle quantum mechanics is equivalent to the quantum theory of a classical field and that a classical field theory mathematically corresponds to the quantum theory of a point particle. In the following subsection it will be performed this iterated quantization procedure.

\subsection{Multiple Quantization and Quantum Theoretical Field Equations}

One begins with a simple binary alternative with two possible values

\begin{equation}
a = \left(\ 1 \ ,\ 2 \ \right).
\end{equation}
If this alternative is quantized, there are assigned complex numbers to the two values of the alternative and thus 
one obtains a Weyl spinor

\begin{equation}
u = \left(\begin{matrix}u_1\\u_2 \end{matrix}\right),
\label{spinor_ur-object}
\end{equation}
being an element of a two dimensional Hilbert space and representing the quantum state $|u \rangle$ of a single ur object.
This means that the probability $p$ to find one of the two values of the binary alternative is the squared inner product between the spinor and the basis spinors

\begin{equation}
u_1=\left(\begin{matrix}1\\0 \end{matrix}\right)\quad,\quad u_2=\left(\begin{matrix}0\\1 \end{matrix}\right)
\label{basis_spinor-space}
\end{equation}
representing the alternative

\begin{equation}
p_u (u_1)=|\langle u_1|u\rangle|^2,\quad p_u (u_2)=|\langle u_2|u\rangle|^2.
\end{equation}
The obtained spinor u can be mapped to a Minkowski vector by using the Pauli matrices the unit matrix in two dimensions included

\begin{equation}
\sigma^0=\left(\begin{matrix}1&0\\ 0&1 \end{matrix}\right),
\sigma^1=\left(\begin{matrix}0&1\\ 1&0 \end{matrix}\right),
\sigma^2=\left(\begin{matrix}0&-i\\ i&0 \end{matrix}\right),
\sigma^3=\left(\begin{matrix}1&0\\ 0&-1 \end{matrix}\right),\\
\end{equation}
according to

\begin{equation}
k^\mu = u^{\dagger} \sigma^\mu u,
\label{Minkowski_vector}
\end{equation}
where $u^{\dagger}$ denotes the adjoint spinor. The obtained vector $k^\mu$ fulfils the relation

\begin{equation}
k_\mu k^\mu=0. 
\label{dispersion_relation}
\end{equation}
A transformation of the spinor u with an element of $SU(2)$ is equivalent to a rotation within 
the subspace corresponding to the components obtained from the Pauli matrices without the unit matrix and thus 
to a $SO(3)$ transformation in a three dimensional real vector space which is identified with physical space
as an ontological assumption. A transformation of the spinor u with an element of $SL(2,\mathbb{C})$ influences all
components of the Minkowski vector but leaves the expression ($\ref{dispersion_relation}$) invariant and therefore 
corresponds to a Lorentz transformation. 
The Minkowski vector obtained in ($\ref{Minkowski_vector}$) or the corresponding spinor of the ur object ($\ref{spinor_ur-object}$) can be regarded as a classical quantity with respect to a further quantization step. 
Such a quantization step can be performed by postulating commutation relations for the spinors describing the ur objects

\begin{equation}
[\hat u_r, \hat u^{\dagger}_s]=\delta_{rs},
\label{commutator_ur-object}
\end{equation} 
where the brackets denote the commutator $[A,B]=AB-BA$ and the indices $r$ and $s$ can take the values one and two. 
Thus the spinor describing an ur object becomes an operator which can be expressed by introduction of creation and annihilation operators referring to single ur objects, denoted by $a_r$ and $a_r^{\dagger}$, where the index r denotes the state of the created ur object and thus also takes the values one and two. The operator of an ur object and its adjoint look as follows

\begin{eqnarray}
\hat u=a_1 u_1+a_2 u_2,\nonumber\\
\hat u^{\dagger}=a_1^{\dagger} u_1^{*}+a_2^{\dagger} u_2^{*},
\end{eqnarray}
whereas $u_1$ and $u_2$ are defined by ($\ref{basis_spinor-space}$) and the operators $a_r$ and $a_r^{\dagger}$ fulfil as usual the following Lie Algebra

\begin{equation}
[a_r,a_s]=0\quad,\quad [a_r^\dagger,a_s^\dagger]=0\quad,\quad [a_r,a_s^\dagger]=\delta_{rs}.
\label{creation_annihilation_ur-objects}
\end{equation}
That the ur objects have to obey Bose statistics is the reason why they obey commutation relations rather than anti commutation relations. If they obeyed Fermi statistics, because of the Pauli principle there could only exist two ur objects in the universe corresponding to the two possible states an ur object can take. One can construct arbitrary states within the tensor space of many ur objects by applying the operators to the vacuum state $|0 \rangle$. Application of an annihilation operator $a_r$ to the vacuum state $|0 \rangle$ gives zero and application of a creation operator to the vacuum state $|0 \rangle$ leads to the state of a single ur object

\begin{equation}
a_r |0 \rangle=0 \quad,\quad a_r^\dagger |0 \rangle=|u_r \rangle.
\end{equation}
The Minkowski vector $k^\mu$ defined by an ur object according to ($\ref{Minkowski_vector}$) also becomes an operator 
$\hat k^\mu$ acting on a quantum state $\varphi(k^\mu)$ which is obtained by assigning complex values to the possible 
values of the Minkowski vector

\begin{equation}
k^\mu \rightarrow \varphi(k^\mu).
\label{transition_vector_wavefunction}
\end{equation}
The transition ($\ref{transition_vector_wavefunction}$) corresponding to this iterated quantization ($\ref{commutator_ur-object}$),($\ref{creation_annihilation_ur-objects}$) can be performed by considering the inner 
product between the eigenstates of the operator $\hat k^\mu$ and a state $|\varphi \rangle$ in the obtained tensor space of ur objects $\varphi(k^\mu)=\langle k^\mu|\varphi \rangle$. Thus one arrives at a wave function which squared inner product describes according to the postulates of quantum theory the probability to get the corresponding value $p(k^\mu)=|\langle k^\mu|\varphi \rangle|^2=|\varphi(k^\mu)|^2$. 
If one holds the ontological assumption that $k^\mu$ represents a momentum in real Minkowski space-time, then the wave function $\varphi(k^\mu)$ can be interpreted as a wave function describing a particle in real space-time. Since the Minkowski vector ($\ref{Minkowski_vector}$) fulfils the algebraic relation

\begin{equation}
k_\mu \sigma^\mu u=0,
\label{relation_k}
\end{equation}
the wave function is only allowed to have values unequal to zero for values of $k^\mu$ fulfilling the constraint ($\ref{relation_k}$). Thus one is led to a constraint on the obtained wave function which can be implemented in 
the sense of Dirac as a condition on the state by applying the operator $\hat k_\mu \sigma^\mu \hat u$ to the state $\varphi(k^\mu)$ and postulating according to ($\ref{relation_k}$) that this expression is equal to zero. 
If one defines a spinor wave function

\begin{equation}
\psi(k^\mu)=u \varphi(k^\mu),
\label{spinor_wavefunction}
\end{equation}
where u describes the ur object of the first quantization step from which the Minkowski vector $k^\mu$ is constructed
according to ($\ref{Minkowski_vector}$), this leads to the following equation

\begin{equation}
k_\mu \sigma^\mu \psi(k^\mu)=0.
\end{equation}
A Fourier transformation introducing a new wave function $\Psi(x^\mu)$ depending on the new coordinates $x^\mu$

\begin{equation}
\Psi(x^\mu)=\int d^4 k \psi(k^\mu) e^{ik_\mu x^\mu}
\label{Fourier}
\end{equation}
leads to the Weyl equation

\begin{equation}
i \sigma^\mu \partial_\mu \Psi(x^\mu)=0.
\label{Weyl_equation}
\end{equation}
The coordinates $x^\mu$ the new wave function depends on which are introduced by the Fourier transformation ($\ref{Fourier}$) fulfil as operators $\hat x^\mu$ together with the momentum operators $\hat k^\mu$ a Heisenberg algebra

\begin{equation}
[\hat x^\mu,\hat k_\nu]=i\delta^{\mu}_{\nu}.
\label{commutator_Heisenberg}
\end{equation}
Thus the coordinates $x^\mu$ can be identified as position vectors in Minkowski space. As usual the corresponding position operators $\hat x^\mu$ act on a state $\varphi(k^\mu)$ as a derivative operator $\hat x^\mu=i\frac{\partial}{\partial k_\mu}$. By performance of the Fourier transformation ($\ref{Fourier}$) there appear four free parameters which are contained in an arbitrary vector $a^\mu$, since all functions $\Psi(x^\mu)$ being related by transformation of the coordinate $x^\mu$ according to

\begin{equation}
x^\mu \rightarrow x^\mu+a^\mu,
\label{translation}
\end{equation}
with $a^\mu$ constant correspond to the same wave function $\psi(k^\mu)$. Thus it seems indeed to be justified to interpret the coordinate $x^\mu$ as a position coordinate in real Minkowski space-time, the corresponding transformation ($\ref{translation}$) as a translation in Minkowski space-time and $k^\mu$ as momentum vector in Minkowski space-time
because it does not have the translation degree of freedom. 

\begin{figure}[h]
\centering
\epsfig{figure=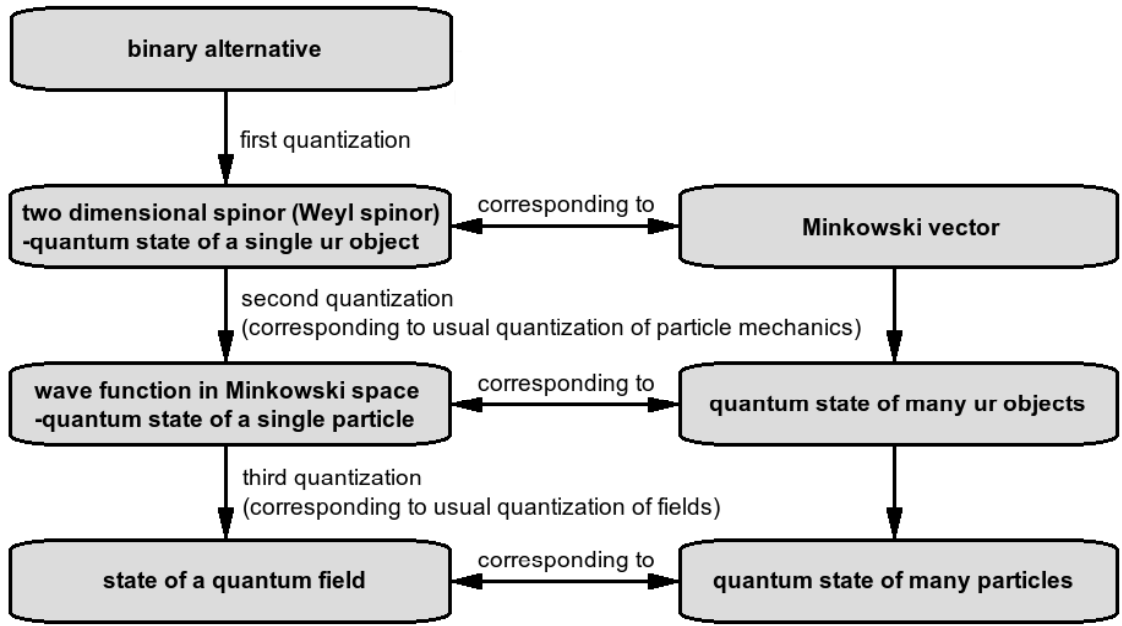,width=15cm}
\caption{\label{multiplequantization} Iteration of quantization of binary alternatives: Beginning from the quantization of a binary alternative one obtains a single ur object being assumed to be the fundamental entity of nature. By performing another quantization one obtains a state containing many ur objects corresponding to a state of a particle. A further quantization of the obtained quantum theoretical wave function in Minkowski space-time leads to usual quantum field theory.}
\end{figure}
This implies that the derived equation ($\ref{Weyl_equation}$) can be interpreted as relativistic wave equation of quantum mechanics describing the dynamics of a quantum state in position space and appearing as a constraint on the second quantization step of a binary alternative in the context of this approach. With respect to this it is important to remember that the quantization of the theory of a single particle leads to quantum mechanics and quantum mechanics is mathematically equivalent to a classical field theory. Multi particle quantum mechanics and quantization of a classical field theory lead to the same theory, namely quantum field theory. Therefore it is possible to interpret field quantization, the so called second quantization, as an iteration of the first quantization of particle mechanics representing in this approach already an iteration of a quantization of a binary alternative assumed to be at the very beginning of physics. Iteration of quantization means to consider a probability distribution of probability distributions. The described iteration of a quantization of binary alternatives and the obtained structures are represented in figure ($\ref{multiplequantization}$).
In this sense a state $\psi(k^\mu)$ obtained by an iterated quantization describes an ensemble of ur objects which are obtained by a quantization of the binary alternative themselves. Since a single ur object corresponds to an exact momentum corresponding to a wave function in position space being completely delocalized one needs of course many ur objects to obtain a wave function $\Psi(x^\mu)$, a quantum state of the second quantization step, which is localized very sharply in position space. 
A second constraint on $\Psi(x^\mu)$ besides $(\ref{Weyl_equation}$) is given by the fact that the overall probability over the spatial part of space-time has to be equal to one. This means that the overall integral over the squared value of the wave function has to be also equal to one

\begin{equation}
\int d^3 x \Psi^{\dagger}(x^\mu) \Psi(x^\mu)=\int d^3 x |\Psi(x^\mu)|^2 = 1.
\end{equation}
Introduction of anti urs corresponding to the two different representations of the Lorentz group in spinor space leads to antiparticles and Dirac spinors and by constructing the momentum vector from two ur objects at the first step of quantization it is possible to obtain momentum vectors for massive particles, to derive the Dirac equation and even to introduce the isospin as internal degree of freedom. The free Maxwell field and its field equations can also be introduced in a similar way \cite{Weizsaecker:1985},\cite{Kober:0910}.\\
Another iteration of the quantization procedure corresponds to a quantization of the spinor wave function $\Psi(x^\mu)$ which is performed as usual by postulating anti commutation relations

\begin{equation}
\left\{\hat \Psi^\alpha(x,t) ,\hat \Psi^{\dagger\beta}(x^\prime,t) \right\}=\delta^3 \left(x-x^\prime \right) \delta^{\alpha \beta},
\label{quantization_spinorfield}
\end{equation}
where $\{A,B\}=AB+BA$ and $x$ denotes the spatial part of the space-time vector the wave function $\Psi(x^\mu)$ depends on. This leads to the customary setting of quantum field theory, where $\Psi(x^\mu)$ itself becomes an operator $\hat \Psi(x^\mu)$ acting on a quantum state $|\Phi \rangle$ of a higher quantization level. As a quantum mechanical state $\Psi(x^\mu)$ contains many ur objects a state $|\Phi \rangle$ where on the level of field quantization $\hat \Psi(x^\mu)$ acts on as an operator contains many particles. Since all particles or fields respectively are constructed from a finite number of ur objects corresponding to a finite number of degrees of freedom, there could be offered the possibility to cure the problem of infinities from the outset which arises in the usual treatment of quantum field theory mandatorily.\\
So far it has been given a description how to obtain quantum theoretical field equations in Minkowski space-time from an iteration of the quantization of a binary alternative assumed to be the fundamental entity of nature as a basic unit of information. One decisive consequence of this approach is the fact that the existence of a position space combined with time to a Minkowski space-time is a derived quality. This is in accordance with the abstract postulates of quantum theory which do not presuppose a position space. Thus there has been derived a space-time manifold in the sense of special relativity. But a more general structure of space-time endowed with a gravitational field according to general relativity is still omitted in this setting. It is the aim of this paper to derive such a richer structure of space-time as a derived property from the quantum theory of ur objects.\\ 
It seems to be possible to obtain such a description if the arbitrariness of the translation vector $a^\mu$ occurring in the above Fourier transformation ($\ref{Fourier}$) is used. All wave functions in position space $\Psi(x^\mu)$ differing only by a translation correspond to the same wave function in momentum space $\psi(k^\mu)$. However, this is only the case as long as the translation vector $a^\mu$ is assumed to be global which means that it is independent of the space-time coordinates $x^\mu$. If it is assumed to be local and thus to be a function of the space-time point $a(x^\mu)$, this assertion does not hold anymore.
To maintain this invariance of the wave function in momentum space the derivative operator has to be modified and thus there has to be introduced a kind of covariant derivative. This means that one arrives at a translation gauge description what will be used to introduce general relativity.\\
To incorporate general relativity there will be considered such space-time dependent translation vectors with respect to the above Fourier transformation ($\ref{Fourier}$) and thus there will be considered the translation gauge theory of gravity in which the non trivial part of the tetrad field itself appears as gauge potential. Based on these considerations it becomes possible to implement gravity in terms of quantum theoretical ur objects. In the context of the quantum theory of ur objects the introduced tetrad field has to be described in an ur theoretic manner, too. This means that the translation gauge field $\theta^m_\mu$ being constructed from ur objects corresponds itself to a wave function (two wave functions, to express it more accurately) representing the second quantization step of an ur alternative. The non trivial part of the tetrad is constructed from two spinor wave functions similar as it is done in \cite{Kober:2009}.
Before formulating a local translation invariant theory in terms of ur objects, there shall be given a short review of the translation gauge theory of gravity and there shall be given reasons why this description represents indeed the adequate gauge theoretical formulation of general relativity.

\section{Translation Gauge Theory of Gravity}

\subsection{Formalism of the Translation Gauge Theory}

In the framework of gauge theories one begins with a free matter field equation being invariant under a certain global transformation group and postulates invariance with respect to the corresponding local transformation group. 
In the translation gauge theory of gravity the Lagrangian of a theory of matter fields is postulated to be invariant
under local space-time translation transformations \cite{Hehl:1976kj},\cite{Lyre:2004B},\cite{Cho:1975dh}. The Lagrangian of a free massless fermionic field reads as follows

\begin{equation}
\mathcal{L}=\bar \Psi i\gamma^\mu \partial_\mu \Psi,
\label{fermion_Lagrangian}
\end{equation}
where the $\gamma^\mu$ denote the Dirac matrices and $\bar \Psi=\Psi^{\dagger} \gamma^0$.
Under a local translation gauge transformation the fermionic field is transformed according to

\begin{equation}
\Psi(x^\mu) \rightarrow T \Psi(x^\mu)=\Psi(x^\mu+a^\mu(x^\mu)),
\label{transformation_matterfield}
\end{equation}
where the translation operator $T$ can be expressed as

\begin{equation}
T=\exp\left[ia^\mu(x^\mu) \hat p_\mu \right]=\exp\left[a^\mu(x^\mu) \partial_\mu \right].
\label{translationoperator}
\end{equation}
The local translations considered in ($\ref{transformation_matterfield}$) and ($\ref{translationoperator}$) 
correspond to diffeomorhisms of the space-time manifold. 
To obtain the Lagrangian corresponding to ($\ref{fermion_Lagrangian}$) being invariant under local gauge transformations one has to replace the derivative $\partial_\mu$ by the covariant derivative $\nabla_\mu$ and thus one has to introduce a field $\theta^\mu_m$ as gauge potential which corresponds to the non trivial part of a tetrad field. The covariant derivative looks as follows

\begin{equation}
\nabla_\mu=\partial_\mu+i\theta_\mu^m \partial_m.
\end{equation}
Under a local translation the tetrad transforms according to

\begin{equation}
\theta^m_\mu \rightarrow T \theta^m_\mu T^{-1}-T^{-1}\partial_\mu T.
\end{equation}
If $a^\mu$ is assumed to be infinitesimal, then this means

\begin{equation}
\theta^m_\mu \rightarrow \theta^m_\mu-\theta^m_\nu\partial_\mu a^\nu.
\end{equation}
Thus the Lagrangian

\begin{equation}
\mathcal{L}=\bar \Psi i\gamma^\mu \left(\partial_\mu+i\theta_\mu^m \partial_m \right)\Psi
=\bar \Psi i\gamma^\mu \nabla_\mu \Psi,
\label{covariant_fermion_Lagrangian}
\end{equation}
is invariant under local translations. The complete tetrad is related to $\theta^m_\mu$ according to

\begin{equation}
e^m_\mu=\delta^m_\mu+\theta^m_\mu,
\label{tetrad_complete}
\end{equation}
and the tetrad field $e^m_\mu$ is itself related to a space-time metric $g_{\mu\nu}$ in the usual way

\begin{equation}
g_{\mu\nu}=e_\mu^m e_{\nu m}.
\label{relation_metric_tetrad}
\end{equation}
Concerning the dynamics of the gravitational field represented by $e_\mu^m$ the decisive quantity is not the curvature but the torsion defined by the tetrad field $e_\mu^m$. It can be seen as the field strength and is given by 

\begin{equation}
T_{\ \mu\nu}^m=\partial_\mu e_\nu^m-\partial_\nu e_\mu^m.
\end{equation}
By defining the torsion with two Minkowski indices

\begin{equation}
T_{\ \ \nu}^{mn}=e^{n\mu} \left(\partial_\mu e_\nu^m-\partial_\nu e_\mu^m \right),
\end{equation}
one can formulate the action in terms of the torsion yielding the dynamics equivalent to the one obtained by the Einstein Hilbert action

\begin{equation}
\mathcal{L}_T=\frac{1}{\kappa^2}\sqrt{-g}\left(\frac{1}{4}T_{\ \mu\nu}^m T_m^{\ \mu\nu}
+\frac{1}{2}T^{mn}_{\ \ \mu} T_{nm}^{\ \ \mu}
-T^m_{\ m\mu}T^{n\ \mu}_{\ n} \right).
\label{Lagrangian_torsion}
\end{equation}
In \cite{Cho:1975dh} there is shown that the action ($\ref{Lagrangian_torsion}$) arises in a natural way from the translation
gauge description of gravity which therefore yields an adequate gauge theory of general relativity.

\subsection{Reasons for the Formulation of Gravity as Translation Gauge Theory}

There are several strong arguments why the translation gauge description of gravity yields indeed the adequate description of gravity.

{\bf 1.} In the theoretical setting of gauge theories in general, the gauge field couples to the conserved Noether current corresponding to the global gauge symmetry group according to Noether's law. The conserved quantity corresponding to the translation group is the energy momentum tensor. And the energy momentum tensor is indeed the quantity the gravitational field is coupled to according to Einstein's field equation. The conserved quantity corresponding to Lorentz transformations is a tensor of third order which is therefore not the correct quantity. 

{\bf 2.} The most general symmetry group of general relativity is the diffeomorphism group. The group of local translations is equivalent to the diffeomorphism group. Since this is the most general symmetry group, it seems to be natural that it should also be the gauge group according to the case of Yang Mills gauge theories.

{\bf 3.} The Lagrangian of Yang Mills gauge theories is quadratic in the field strength. Torsion is the field strength of the translation gauge theory and the Lagrangian arising from this description and being dynamically equivalent to the Einstein Hilbert Lagrangian is quadratic in the field strengths ($\ref{Lagrangian_torsion}$).

{\bf 4.} The gauge field is itself the tetrad field which yields a more appropriate treatment of general relativity than the metric field. This has another three reasons:\\
a) The general relativistic formulation of a fermionic action can only be formulated in terms of a tetrad.\\
b) The tetrad formulation is used in modern approaches to Quantum Gravity.\\
c) The tetrad reflects the real nature of the gravitational field in a more clean way since it directly expresses a kind of relation between coordinates.\\

Therefore the translation group should be preferred as gauge group of general relativity with respect to the SO(3,1), the Lorentz group, for example. The reasons 1-3 for the formulation of gravity as translation gauge theory can be found in \cite{Lyre:2004B} and the justifications a-c of reason 4 can be found in \cite{Rovelli:2004}. Since within the presented approach of a derivation of general relativity within the quantum theory of ur objects the translation gauge description of gravity appears in a natural way, the tetrad field appearing as gauge potential in this description can directly be constructed from spinors and there exist weighty reasons for a description of gravity as translation gauge theory, this approach seems to obtain a strong justification.

\section{Incorporation of General Relativity to the Quantum Theory of Ur Objects}

\subsection{Local Translations and the Tetrad Field expressed by Ur Objects}

The quantum theory of ur objects shall now be considered again. It has already been mentioned that the wave function obtained by the Fourier transformation ($\ref{Fourier}$) has a translation gauge degree of freedom

\begin{equation}
\psi(x^\mu)=\int d^4 k \psi(k^\mu) e^{ik_\mu(x^\mu+a^\mu)}.
\end{equation}
If the translation gauge parameter $a^\mu$ is assumed to depend on the space-time coordinates

\begin{equation}
x^\mu \rightarrow x^\mu+a^\mu(x^\mu),
\end{equation}
then the derivative has to be transformed according to

\begin{equation}
\partial_\mu \rightarrow \partial_\mu-\partial_\mu a^m \partial_m \equiv \partial_\mu+i \theta_\mu^m \partial_m = \nabla_\mu
\label{derivative_localtranslation}
\end{equation}
to obtain still the correct momenta $k^\mu$ and to maintain Heisenberg's commutation relation ($\ref{commutator_Heisenberg}$).
The quantity $\partial_\mu a^\nu$ corresponds to the non trivial part of a tetrad multiplied with a factor i and thus the new derivative in ($\ref{derivative_localtranslation}$) corresponds to the derivative of a translation gauge theory.

If one now wants to interpret the tetrad field as a real physical field and not just in the trivial way as a choice of new coordinates, it has to be the consequence of a representation by ur objects, too. In \cite{Lyre:1996ep} there is constructed a tetrad directly from two ur objects on the first step of quantization, call them u and v,

\begin{eqnarray}
e_0^m=\frac{1}{\sqrt{2}}\left(u^{\dagger} \sigma^m u+v^{\dagger} \sigma^m v\right)
\quad,\quad e_1^m=\frac{1}{\sqrt{2}}\left(v^{\dagger} \sigma^m u+u^{\dagger} \sigma^m v\right), \nonumber\\
e_2^m=\frac{1}{\sqrt{2}}i\left(v^{\dagger} \sigma^m u-u^{\dagger} \sigma^m v\right)
\quad,\quad e_3^m=\frac{1}{\sqrt{2}}\left(u^{\dagger} \sigma^m u-v^{\dagger} \sigma^m v\right).
\label{tetrad_Minkowski}
\end{eqnarray}
This is mathematically in accordance with the tetrad of vectors constructed from two spinors within the twistor theory of Roger Penrose \cite{Penrose:1960eq},\cite{Penrose:1977in},\cite{Penrose:1985jw},\cite{Penrose:1986ca},\cite{Stewart:1990uf}.
Within the twistor approach there is also considered the correspondence between a Minkowski vector and a Weyl spinor. 
A structure yielding a representation of the Minkowski metric within the spinor space which raises and lowers indices is obtained by introduction of the symplectic structure [$\cdot$,$\cdot$] obeying $[\varphi,\chi]=-[\chi,\varphi]$ which can be represented by the matrix $\epsilon_{\alpha\beta}$

\begin{equation}
\epsilon_{\alpha\beta}=\begin{pmatrix} 0 & 1 \\ -1 & 0 \end{pmatrix}
\end{equation}
according to

\begin{equation}
[\varphi,\chi]=\epsilon_{\alpha\beta}\varphi^\alpha \chi^\beta.
\end{equation}
This means that raising and lowering a spinor index with $\epsilon_{\alpha\beta}$ corresponds to an inversion of the spatial coordinates. However, the construction ($\ref{tetrad_Minkowski}$) leads to a tetrad corresponding to a metric which is proportional to the Minkowski metric. In order to obtain general metric fields one has to introduce an extended description of the tetrad according to \cite{Kober:2009}, where there is added a term to the tetrad ($\ref{tetrad_Minkowski}$) which contains derivatives of a couple of spinor fields. The non trivial part $\theta_\mu^m$ of the tetrad field corresponding to the gauge potential of the translation gauge degree of freedom has to be constructed in such a way. This means that one has to construct two spinor wave functions representing already quantum states of the second step of quantization in the setting of a description by ur objects to obtain an adequate tetrad field leading to general metric fields. This is performed in accordance with the construction of the matter field above ($\ref{Minkowski_vector}$),($\ref{transition_vector_wavefunction}$),
($\ref{spinor_wavefunction}$),($\ref{Fourier}$). 
Therefore one has to begin from two free ur objects u and v being independent of the ones within the wave equation describing the state of the matter field. They lead to a couple of Minkowski vectors, call them $k^\mu$ and $l^\mu$

\begin{equation}
k^\mu=u^{\dagger} \sigma^\mu u \quad,\quad l^\mu=v^{\dagger} \sigma^\mu v,
\end{equation}
on the first level of quantization. One the level of the second step of quantization one obtains wave functions, 
call them $\phi$ and $\omega$

\begin{equation}
k^\mu \rightarrow \phi(k^\mu) \quad,\quad l^\mu \rightarrow \omega(l^\mu),
\end{equation}
which contain many ur objects and the quantum constraints lead to the following equations

\begin{eqnarray}
k_\mu \sigma^\mu u\phi(k^\mu)&=&0,\nonumber\\
l_\mu \sigma^\mu v\omega(l^\mu)&=&0.
\end{eqnarray}
Defining the spinor wave functions

\begin{equation}
\varphi(k^\mu)=u \phi(k^\mu) \quad,\quad \chi(l^\mu)=v \omega(m^\mu),
\label{spinor_fields}
\end{equation}
and performing a Fourier transformation

\begin{eqnarray}
\varphi(x^\mu)=\int d^4 k \varphi(k^\mu) e^{i k_\mu x^\mu},\nonumber\\
\chi(x^\mu)=\int d^4 k \chi(k^\mu) e^{i k_\mu x^\mu},
\label{Fourier_varphi_chi}
\end{eqnarray}
leads to the dynamical equations in position space, the Weyl equations

\begin{eqnarray}
i\sigma^\mu \partial_\mu \varphi(x^\mu)=0,\nonumber\\
i\sigma^\mu \partial_\mu \chi(x^\mu)=0.
\label{constraints}
\end{eqnarray}
(It is important to mention that this description represents an approximation, since the gauge degree of freedom from which
the tetrad field arises, is omitted within the description of the spinor fields according to ($\ref{constraints}$).)\\
It is now possible to construct the non trivial part of the tetrad field from the above quantities

\begin{equation}\normalfont
\theta^m_\mu = i\left(\varphi^{\dagger} \sigma^m \partial_\mu \varphi+\chi^{\dagger} \sigma^m \partial_\mu \chi \right).
\label{tetrad_dynamical}
\end{equation}
According to ($\ref{constraints}$) the tetrad field obeys the constraint

\begin{equation}
\partial_m \theta^m_\mu=0.
\label{constraint_tetrad}
\end{equation}
It is also possible to define a Dirac spinor field $\psi_D$ 

\begin{equation}
\psi_D=\left(\begin{matrix}\varphi\\ i \sigma^2 \chi^{*}\end{matrix}\right)
\label{Dirac_spinor}
\end{equation}
and to write the non trivial part of the tetrad as follows

\begin{equation}
\theta^m_\mu=i\bar \psi_D \gamma^m \partial_\mu \psi_D.
\label{Dirac_spinor_tetrad}
\end{equation}
The obtained tetrad field ($\ref{tetrad_dynamical}$) does only represent the non trivial part of the tetrad field $e_\mu^m$. Since the tetrad must not vanish but has to lead to the Minkowski metric ($\ref{relation_metric_tetrad}$) for constant spinor fields $\varphi$ and $\chi$, there has to be added the term being defined by the ur objects of the first quantization step ($\ref{tetrad_Minkowski}$), u and v, to ($\ref{tetrad_dynamical}$) and thus the complete tetrad has to look as follows

\begin{equation}
e_\mu^m = \frac{1}{2}
\left(\begin{matrix}
u^{\dagger} \sigma^m u+v^{\dagger} \sigma^m v\\
u^{\dagger} \sigma^m v+v^{\dagger} \sigma^m u\\
iu^{\dagger} \sigma^m v-iv^{\dagger} \sigma^m u\\
u^{\dagger} \sigma^m u-v^{\dagger} \sigma^m v
\end{matrix}\right)
+i\left(\varphi^{\dagger} \sigma^m \partial_\mu \varphi+\chi^{\dagger} \sigma^m \partial_\mu \chi \right).
\label{tetrad_complete}
\end{equation}

\subsection{Dynamics of a Matter Field interacting with the Gravitational Field}

To formulate the complete dynamics of a matter field constructed from ur objects according to ($\ref{Minkowski_vector}$),
($\ref{transition_vector_wavefunction}$),($\ref{spinor_wavefunction}$),($\ref{Fourier}$) which incorporates the interaction with the gravitational field expressed by ur objects, there has to be introduced a spin connection which has to be derived from the tetrad in the usual way and depends therefore on the spinor wave functions $\varphi$ and $\chi$ from which the gravitational field as tetrad field is built. The connection looks as follows

\begin{equation}
A_\mu^{\alpha\beta}(\varphi,\chi)=2e^{\nu[\alpha}(\varphi,\chi) \partial_{[\mu} e_{\nu]}^{\beta]}(\varphi,\chi)+e_{\mu\rho}(\varphi,\chi)e^{\nu\alpha}(\varphi,\chi)e^{\sigma\beta}(\varphi,\chi)\partial_{[\sigma} e_{\nu]}^{\rho}(\varphi,\chi),
\end{equation}
where the brackets $[\cdot \cdot]$ denote antisymmetrisation with respect to the corresponding indices.
Thus the field equation of the free matter field derived from the quantum theory of ur objects under incorporation of the local translation degree of freedom expressed in terms of ur objects by itself and corresponding to a general relativistic setting looks as follows

\begin{equation}
i\sigma^\mu \left(\partial_\mu+i\theta^m_\mu(\varphi,\chi)\partial_m+iA_\mu^{\alpha\beta}(\varphi,\chi)\Sigma_{\alpha\beta}\right)\Psi \equiv i\sigma^\mu D_\mu \Psi=0,
\label{equation_matterfield-gravity}
\end{equation}
where the $\Sigma_{\alpha\beta}$ are the generators of the Lorentz group in the spinor representation 

\begin{equation}
\Sigma_{\alpha\beta}=\frac{i}{2}[\sigma_\alpha,\sigma_\beta],
\end{equation}
fulfilling the corresponding commutation relations

\begin{equation}
[\Sigma_{\alpha\beta},\Sigma_{\gamma\delta}]=\eta_{\gamma\beta}\Sigma_{\alpha\delta}-\eta_{\gamma\alpha}
\Sigma_{\beta\delta}+\eta_{\gamma\beta}\Sigma_{\delta\alpha}-\eta_{\delta\alpha}\Sigma_{\gamma\beta}.
\end{equation}
If the spin connection is expressed elaborately in terms of the tetrad field, equation ($\ref{equation_matterfield-gravity}$)
looks as follows

\begin{equation}
i\sigma^\mu \left(\partial_\mu+i\theta_\mu^m(\varphi,\chi) \partial_m
+i\left(2e^{\nu[\alpha}(\varphi,\chi) \partial_{[\mu} e_{\nu]}^{\beta]}(\varphi,\chi)
+e_{\mu\rho}(\varphi,\chi)e^{\nu\alpha}(\varphi,\chi)e^{\sigma\beta}(\varphi,\chi)\partial_{[\sigma} e_{\nu]}^{\rho}(\varphi,\chi)\right)\Sigma_{\alpha\beta}\right)\Psi=0.
\end{equation}
This fundamental field equation of matter corresponds to the following action

\begin{equation}
S_{matter}=\int d^4 x \det[e_\mu^m(\varphi,\chi)]\Psi^{\dagger} i\sigma^\mu \left(\partial_\mu+i\theta_\mu^m(\varphi,\chi)\partial_m
+iA_\mu^{\alpha\beta}(\varphi,\chi)\Sigma_{\alpha\beta}\right)\Psi. 
\end{equation}

\subsection{Dynamics of the Gravitational Field}

The above incorporation of gravity is not exact. This is because of the fact that the two spinor fields $\varphi$ and $\chi$ constituting the tetrad field have to interact with themselves because they are also constructed from the iteration of the quantization procedure of the binary alternatives and thus they also have the translation gauge invariance. A more exact consideration of the dynamics of the spinor fields is obtained by formulating the field equations for these fields incorporating this self interaction. This leads to the following equations

\begin{eqnarray}
i\sigma^\mu \left(\partial_\mu+e^m_\mu(\varphi,\chi) \partial_m+iA_\mu^{\alpha\beta}(\varphi,\chi)\Sigma_{\alpha\beta} \right)\varphi=
i\sigma^\mu D_\mu \varphi=0,\nonumber\\
i\sigma^\mu \left(\partial_\mu+e^m_\mu(\varphi,\chi) \partial_m+iA_\mu^{\alpha\beta}(\varphi,\chi)\Sigma_{\alpha\beta} \right)\chi=
i\sigma^\mu D_\mu \chi=0,
\label{dynamics_gravity_spinor}
\end{eqnarray}
corresponding to the action

\begin{equation}
S_{gravity}=\int d^4 x \det[e_\mu^m(\varphi,\chi)]\left(\varphi^{\dagger}i\sigma^\mu D_\mu \varphi\right.\nonumber\\
\left.+\chi^{\dagger}i\sigma^\mu D_\mu \chi\right).
\label{gravity_action}
\end{equation}
The constraint ($\ref{constraint_tetrad}$) reads accordingly

\begin{equation}
D_m e^m_\mu=0,
\label{constraint_tetrad_exact}
\end{equation}
which corresponds to the Gauss constraint in Ashtekar's formulation of canonical general relativity
and the dynamics of the tetrad field $e_\mu^m(\varphi,\chi)$ and the corresponding metric field $g_{\mu\nu}(\varphi,\chi)$ respectively is determined by the dynamics of the spinor fields ($\ref{dynamics_gravity_spinor}$). This dynamics should have a dynamics of the tetrad field as consequence which corresponds to the gravity action ($\ref{Lagrangian_torsion}$) of the usual translation gauge description of gravity being equivalent to the Einstein Hilbert action. At least it should contain it as a certain approximation. However, to relate this two different descriptions of the dynamics of gravity represents a very difficult mathematical task. Therefore concerning the derivation of the dynamics of the tetrad field from the one of the spinor fields ($\ref{dynamics_gravity_spinor}$) it will just be considered the linear approximation without self interaction of the spinor fields $\varphi$ and $\chi$ meanwhile. 
In the linear approximation equations ($\ref{dynamics_gravity_spinor}$) convert to ($\ref{constraints}$). In this approximation the tetrad field $e_\mu^m(\varphi,\chi)$ obeys the following wave equation

\begin{equation}
\partial_\rho \partial^\rho e_\mu^m(\varphi,\chi)=i\left(2\partial_\rho \varphi^{\dagger} \sigma^m \partial_\mu \partial^\rho \varphi+2\partial_\rho \chi^{\dagger} \sigma^m \partial_\mu \partial^\rho \chi \right),
\label{wave_equation_tetrad}
\end{equation}
whereas the term on the right hand side vanishes if ($\ref{wave_equation_tetrad}$) is integrated over the whole space

\begin{equation}
\int d^3 x \partial_\rho \partial^\rho e_\mu^m(\varphi,\chi)=0.
\end{equation}
This can be seen as follows: Since the spinor fields obey Weyl equations ($\ref{constraints}$), both components of the
spinor fields obey the wave equation

\begin{eqnarray}
\partial_\mu \partial^\mu \varphi=0,\nonumber\\
\partial_\mu \partial^\mu \chi=0.
\end{eqnarray}
Applying the operator $\partial_\mu \partial^\mu$ to the tetrad field $e_\mu^m(\varphi,\chi)$ leads to

\begin{eqnarray}
&&i\partial_\rho \partial^\rho \left(\varphi^{\dagger} \sigma^m \partial_\mu \varphi+\chi^{\dagger} \sigma^m \partial_\mu \chi \right),\nonumber\\
&=&i\left(\partial_\rho \partial^\rho \varphi^{\dagger} \sigma^m \partial_\mu \varphi+\varphi^{\dagger} \sigma^m \partial_\mu \partial_\rho \partial^\rho \varphi
+\partial_\rho \partial^\rho \chi^{\dagger} \sigma^m \partial_\mu \chi+\chi^{\dagger} \sigma^m \partial_\mu \partial_\rho \partial^\rho \chi \right.\nonumber\\
&&\left.+2\partial_\rho \varphi^{\dagger} \sigma^m \partial_\mu \partial^\rho \varphi+2\partial_\rho \chi^{\dagger} \sigma^m \partial_\mu \partial^\rho \chi \right)\nonumber\\
&=&i\left(2\partial_\rho \varphi^{\dagger} \sigma^m \partial_\mu \partial^\rho \varphi+2\partial_\rho \chi^{\dagger} \sigma^m \partial_\mu \partial^\rho \chi \right).
\end{eqnarray}
The remaining term $i\left(2\partial_\rho \varphi^{\dagger} \sigma^m \partial_\mu \partial^\rho \varphi+2\partial_\rho \chi^{\dagger} \sigma^m \partial_\mu \partial^\rho \chi \right)$
vanishes if it is integrated over the whole space since it holds $\int d^3 x \varphi^{\dagger} \varphi < \infty$ and $\int d^3 x \chi^{\dagger} \chi < \infty$ and therefore $\varphi$ and $\chi$ vanish at infinity at the level of first quantization or second quantization, respectively, if one begins the numbering at the binary alternative. This can be seen by partial integration of the remaining expression

\begin{equation}
\int d^3 x \left(2\partial_\rho \varphi^{\dagger} \sigma^m \partial_\mu \partial^\rho \varphi+2\partial_\rho \chi^{\dagger} \sigma^m \partial_\mu \partial^\rho \chi \right)
=-\int d^3 x \left(2\varphi^{\dagger} \sigma^m \partial_\mu \partial_\rho \partial^\rho \varphi+2\chi^{\dagger} \sigma^m \partial_\mu \partial_\rho \partial^\rho \chi \right)=0.
\end{equation}
Thus the usual free wave equation is valid for the tetrad if it is integrated over the whole space ($\ref{wave_equation_tetrad}$) and describes the dynamics of the tetrad within the linear approximation 
besides the constraint ($\ref{constraint_tetrad_exact}$).
With respect to the linear approximation the metric field is according to ($\ref{relation_metric_tetrad}$) and ($\ref{tetrad_complete}$) given by

\begin{equation}
g_{\mu\nu}(\varphi,\chi)=\eta_{\mu\nu}+\theta_{\mu\nu}(\varphi,\chi)+\theta_{\nu\mu}(\varphi,\chi)+\mathcal{O}(\theta^2)
\end{equation}
And this implies that the metric field $g_{\mu\nu}(\varphi,\chi)$ also obeys the usual free wave equation if it is integrated over the whole space 

\begin{equation}
\int d^3 x \partial_\rho \partial^\rho g_{\mu\nu}(\varphi,\chi)=0.
\end{equation}
Thus the linear approximation of the obtained gravity theory corresponds to the linear approximation of usual general relativity, if it is integrated over the whole space. This means that it has been constructed a formulation of gravity expressed by ur objects which corresponds to the translation gauge theory of gravity at least approximatively whereas the translation gauge degree of freedom arises from the ur hypothesis itself.

\subsection{Quantum Description of the Gravitational Field}

Since the gravitational field, like matter fields, is described by spinor fields already representing a state of many ur objects corresponding to the second step of quantization, there is obtained a quantum description of the tetrad field by performing another quantization, a quantization of the spinor fields $\varphi$ and $\chi$, corresponding to a field quantization of a spinor field according to ($\ref{quantization_spinorfield}$). There can be used the quantization procedure 
of usual quantum field theory from now on.  
The fields $\varphi$ and $\chi$ whose dynamics is described by ($\ref{dynamics_gravity_spinor}$)
have to be quantized according to a field theory of fermionic fields. 
The quantization rules are given by the following anti commutation relations

\begin{eqnarray}
\left\{\hat\varphi^\alpha(x,t),\hat \Pi_\varphi^\beta(x^\prime,t) \right\}=i\delta^{\alpha\beta}\delta^3(x-x^\prime)
\quad,\quad\left\{\hat \varphi^\alpha(x,t),\hat \varphi^\beta(x^\prime,t) \right\}=\left\{\hat \Pi_\varphi^\alpha(x,t),\hat \Pi_\varphi^\beta(x^\prime,t) \right\}=0,\nonumber\\
\left\{\hat \chi^\alpha(x,t),\hat \Pi_\chi^\beta(x^\prime,t) \right\}=i\delta^{\alpha\beta}\delta^3(x-x^\prime)
\quad,\quad\left\{\hat \chi^\alpha(x,t), \hat \chi^\beta(x^\prime,t) \right\}=\left\{\hat \Pi_\chi^\alpha(x,t),\hat \Pi_\chi^\beta(x^\prime,t) \right\}=0,\nonumber\\
\left\{\hat \varphi^\alpha(x,t),\hat \chi^\beta(x^\prime,t) \right\}=
\left\{\hat \varphi^\alpha(x,t),\hat \Pi_\chi^\beta(x^\prime,t) \right\}=
\left\{\hat \Pi_\varphi^\alpha(x,t),\hat \chi^\beta(x^\prime,t) \right\}=
\left\{\hat \Pi_\varphi^\alpha(x,t),\hat \Pi_\chi^\beta(x^\prime,t) \right\}=0,\nonumber\\
\label{anticommutation_relation}
\end{eqnarray}
where $x$ denotes again the spatial part of the space-time vector the wave functions $\varphi$ and $\chi$ depend on
and $\Pi_\varphi$ and $\Pi_\chi$ denote their canonical momenta. In a linear approximation neglecting the self coupling of the fields contained in ($\ref{dynamics_gravity_spinor}$), the canonical momenta are defined by the action corresponding to ($\ref{constraints}$), which is the linearized version of ($\ref{gravity_action}$)

\begin{equation}
\Pi_\varphi=i\varphi^{\dagger} \quad,\quad 
%=\frac{\delta S_{gravity}}{\delta \partial_0 \varphi}
\Pi_\chi=i\chi^{\dagger}.
%=\frac{\delta S_{gravity}}{\delta \partial_0 \chi}
\end{equation}
It has again to be mentioned that this quantization approach represents the third step of the quantization of the two binary alternatives leading to the ur objects u and v one the first level of quantization. The anti commutation relations ($\ref{anticommutation_relation}$) of the spinor fields imply non trivial anti commutation relations and corresponding commutation relations for the tetrad field

\begin{equation}
[\hat e_\mu^m(\varphi,\chi),\hat e_\nu^n(\varphi,\chi)] \neq 0,
\end{equation}
depending on the anti commutation relations of the spinor fields ($\ref{anticommutation_relation}$) and thus there is given a quantum theoretical behaviour for the gravitational field which is derived from the one of the spinor fields.
If the spinor field operators are expressed in the usual way by introduction of creation and annihilation operators, they look as follows

\begin{eqnarray} 
\hat \varphi(x,t)=\sum_{\pm \sigma} \frac{1}{N}\int d^3 k a_{\varphi}(k,\sigma)u(k,\sigma) e^{-ik_\mu x^\mu},\nonumber\\
\hat \chi(x,t)=\sum_{\pm \sigma} \frac{1}{N}\int d^3 k a_{\chi}(k,\sigma)u(k,\sigma) e^{-ik_\mu x^\mu},
\label{spinorfield_operators}
\end{eqnarray}
where N is a normalization factor, $\sigma$ refers to the spin and the creation and annihilation operators $a_{\varphi}(k,\sigma)$, $a^{\dagger}_{\varphi} (k,\sigma)$ and $a_{\chi}(k,\sigma)$, $a^{\dagger}_{\chi}(k,\sigma)$ obey anti commutation relations

\begin{eqnarray}
\left\{a_{\varphi}(k,\sigma),a^{\dagger}_{\varphi}(k^{\prime},\sigma^{\prime})\right\}
=\delta_{\sigma,\sigma^{\prime}}\delta(k-k^{\prime})\quad,\quad
\left\{a_{\varphi}(k,\sigma),a_{\varphi}(k^{\prime},\sigma^{\prime})\right\}=
\left\{a^{\dagger}_{\varphi}(k,\sigma),a^{\dagger}_{\varphi}(k^{\prime},\sigma^{\prime})\right\}=0,\nonumber\\
\left\{a_{\chi}(k,\sigma),a^{\dagger}_{\chi}(k^{\prime},\sigma^{\prime})\right\}
=\delta_{\sigma,\sigma^{\prime}}\delta(k-k^{\prime})\quad,\quad
\left\{a_{\chi}(k,\sigma),a_{\chi}(k^{\prime},\sigma^{\prime})\right\}=
\left\{a^{\dagger}_{\chi}(k,\sigma),a^{\dagger}_{\chi}(k^{\prime},\sigma^{\prime})\right\}=0,\nonumber\\
\left\{a_{\varphi}(k,\sigma),a_{\chi}(k^{\prime},\sigma^{\prime})\right\}=
\left\{a_{\varphi}(k,\sigma),a^{\dagger}_{\chi}(k^{\prime},\sigma^{\prime})\right\}=
\left\{a_{\chi}(k,\sigma),a^{\dagger}_{\varphi}(k^{\prime},\sigma^{\prime})\right\}=
\left\{a^{\dagger}_{\varphi}(k,\sigma),a^{\dagger}_{\chi}(k^{\prime},\sigma^{\prime})\right\}=0.
\label{creation_annihilation_tetrad}
\end{eqnarray}
(These operators are not to be confused with the creation and annihilation operators in ($\ref{creation_annihilation_ur-objects}$), which refer directly to single ur objects, whereas the operators defined 
here refer to a higher quantization level. Of course it has in principle to be possible to express the introduced field operators also in terms of creation and annihilation operators referring directly to the tensor space of single ur objects).\\
Introduction of the operators $\hat \varphi$ and $\hat \chi$ defines a Hilbert space of quantum states $|\Phi(\varphi,\chi)\rangle$ over the state space of the fields $\varphi$ and $\chi$.
In analogy to ($\ref{Dirac_spinor}$) and ($\ref{Dirac_spinor_tetrad}$) it can be given an equivalent description by defining a Dirac operator

\begin{equation}
\hat \psi_D=\left(\begin{matrix}\hat \varphi\\i\sigma^2\hat \chi^{*} \end{matrix}\right),
\end{equation}
and postulating the anti commutation relation

\begin{equation}  
\{\hat \psi^{\alpha}_D(x,t), \hat \psi^{\dagger\beta}_{D}(x^{\prime},t)\}=\delta^{\alpha\beta}\delta(x-x^{\prime}).
\label{commutator_Dirac}
\end{equation}  
By defining the operators $b(k,\sigma)$ and $d(k,\sigma)$ according to

\begin{equation}
b(k,\sigma)=a_\varphi(k,\sigma) \quad,\quad d(k,\sigma)=a_\chi^{\dagger}(k,\sigma),
\label{definition_bd}
\end{equation}
there is obtained the usual form of a Dirac field operator and its adjoint from which the tetrad field as gravitational field is derived

\begin{eqnarray} 
\hat \psi_D(x,t)=\sum_{\pm \sigma} \frac{1}{N}\int d^3 k \left(b(k,\sigma)u(k,\sigma) e^{-ik_\mu x^\mu}+d^{\dagger}(k,\sigma)v(k,\sigma) e^{ik_\mu x^\mu}\right),\nonumber\\
\hat \psi_D^{\dagger}(x,t)=\sum_{\pm \sigma} \frac{1}{N}\int d^3 k \left(b^{\dagger}(k,\sigma)u^{\dagger}(k,\sigma) e^{ik_\mu x^\mu}+d(k,\sigma)v^{\dagger}(k,\sigma) e^{-ik_\mu x^\mu}\right),
\label{spinorfield_operator_Dirac}
\end{eqnarray}
where $v$ describes the Majorana conjugated spinor of $u$.
The tetrad field operator is obtained by replacing the spinors in ($\ref{tetrad_dynamical}$) through the corresponding operators defined by ($\ref{anticommutation_relation}$) 

\begin{equation} 
\hat \theta^m_\mu(x,t)=i\left(\hat \varphi^{\dagger}(x,t) \sigma^m \partial_\mu \hat \varphi(x,t)+\hat \chi^{\dagger}(x,t) \sigma^m \partial_\mu \hat \chi(x,t) \right),
\end{equation}
or by replacing the Dirac spinor in ($\ref{Dirac_spinor_tetrad}$) by the corresponding operator defined by ($\ref{commutator_Dirac}$) respectively.
By inserting ($\ref{spinorfield_operators}$) to ($\ref{tetrad_dynamical}$) or ($\ref{tetrad_complete}$) respectively, there is obtained the operator of the tetrad field expressed elaborately by the creation and annihilation operators  

\begin{eqnarray}
\hat \theta^m_\mu(x,t)=\frac{1}{N^2}\left(\sum_{\pm \sigma^{\prime}} \int d^3 k^{\prime}
a^{\dagger}_{\varphi}(k^{\prime},\sigma^{\prime})u^{\dagger}(k^{\prime},\sigma^{\prime})e^{ik_\mu^{\prime} x^\mu}
\sigma^m \sum_{\pm \sigma} \int d^3 k a_{\varphi}(k,\sigma)u(k,\sigma)i\partial_\mu e^{-ik_\mu x^\mu}\right.\nonumber\\
\left.+\sum_{\pm \sigma^{\prime}} \int d^3 k^{\prime}a^{\dagger}_{\chi}(k^{\prime},\sigma^{\prime})v(k^{\prime},\sigma^{\prime})
e^{ik_\mu^{\prime} x^\mu}\sigma^m \sum_{\pm \sigma} \int d^3 k a_{\chi}(k,\sigma)v^{\dagger}
(k,\sigma)i\partial_\mu e^{-ik_\mu x^\mu}\right).
\end{eqnarray}
This can be transformed to

\begin{eqnarray}
\hat \theta^m_\mu(x,t)=\sum_{\pm \sigma^{\prime} \pm \sigma} \frac{1}{N^2}\int\int d^3 k^{\prime} d^3 k
\left(a^{\dagger}_{\varphi}(k^{\prime},\sigma^{\prime})a_{\varphi}(k,\sigma)u^{\dagger}(k^{\prime},\sigma^{\prime}) \sigma^m u(k,\sigma)k_\mu e^{i(k_\mu^{\prime}-k_\mu)x^\mu}\right.\nonumber\\
\left.+a^{\dagger}_{\chi}(k^{\prime},\sigma^{\prime})a_{\chi}(k,\sigma)v(k^{\prime},\sigma^{\prime}) \sigma^m v^{\dagger}(k,\sigma)k_\mu e^{i(k_\mu^{\prime}-k_\mu)x^\mu}\right),
\label{tetrad_operator_Weyl}
\end{eqnarray}
or equivalently to

\begin{eqnarray}
\hat \theta^m_\mu(x,t)=\sum_{\pm \sigma^{\prime} \pm \sigma} \frac{1}{N^2}\int\int d^3 k^{\prime} d^3 k
\left(b^{\dagger}(k^{\prime},\sigma^{\prime})b(k,\sigma)u^{\dagger}(k^{\prime},\sigma^{\prime}) \sigma^m u(k,\sigma)k_\mu e^{i(k_\mu^{\prime}-k_\mu)x^\mu}\right.\nonumber\\
\left.+d(k^{\prime},\sigma^{\prime})d^{\dagger}(k,\sigma)v(k^{\prime},\sigma^{\prime}) \sigma^m v^{\dagger}(k,\sigma)k_\mu e^{i(k_\mu^{\prime}-k_\mu)x^\mu}\right),
\label{tetrad_operator_Dirac}
\end{eqnarray}
if the definition ($\ref{definition_bd}$) is used. ($\ref{tetrad_operator_Weyl}$) and ($\ref{tetrad_operator_Dirac}$) represent the quantum description of the gravitational field within the presented approach. There remains the open question whether the considered iteration of quantization has to be continued and what the meaning of such a description would be.
%It shall just be mentioned that the next step of quantization could lead to a kind of quantum cosmology. But this is just an assumption.

\subsection{Conceptual Issues concerning the Nature of Space-Time}

\begin{figure}[h]
\centering
\epsfig{figure=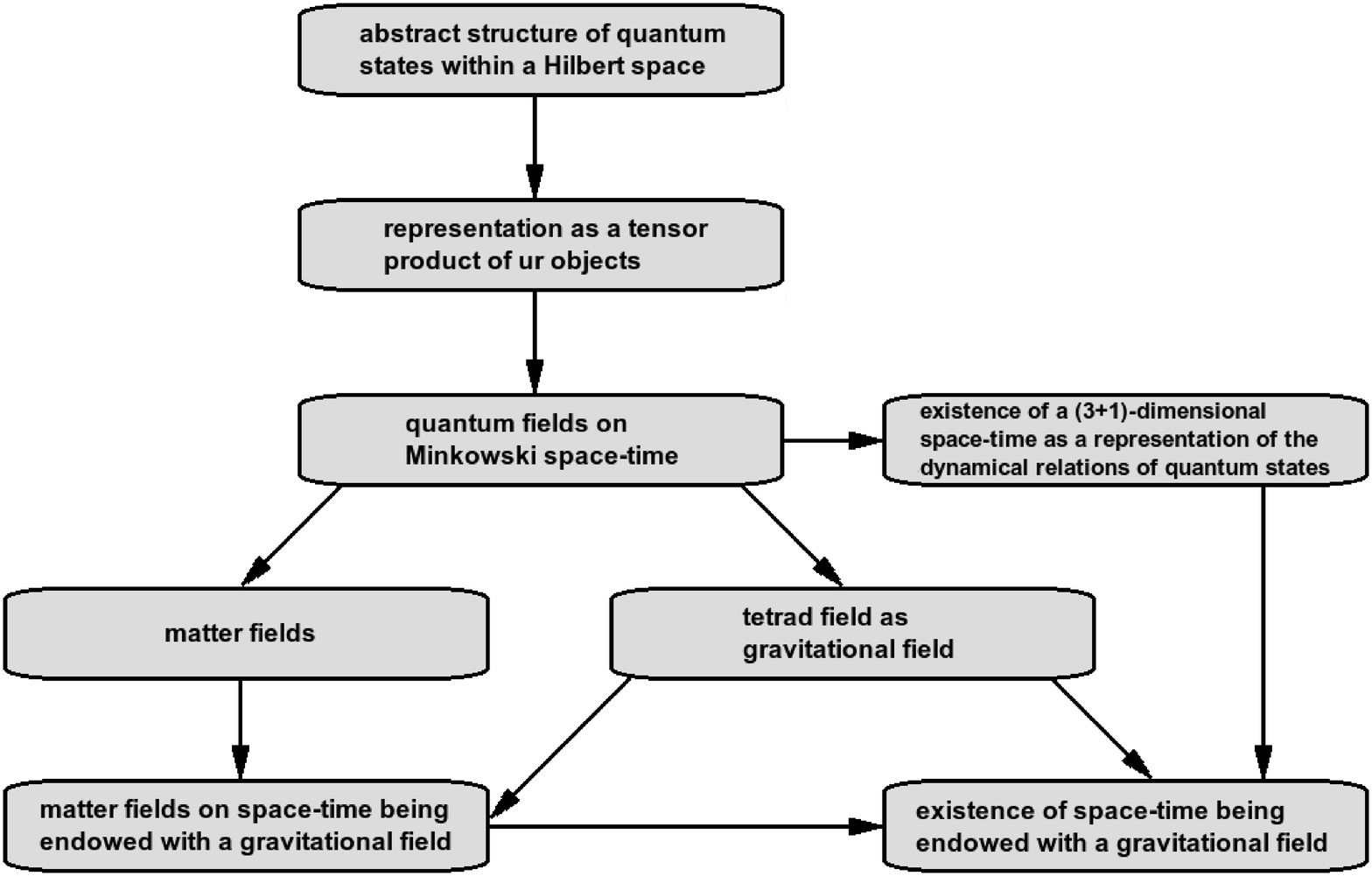, width=19cm}
\caption{\label{spacetimequantumstates} The hierarchy of the entities existing in nature according to the presented approach:
The basic entities of nature are quantum states interpreted as abstract information. They can be thought to be composed by ur objects which symmetry corresponds to the symmetry of Minkowski space-time. This gives rise to the representation of the dynamical relations of quantum states corresponding to usual space-time. The gravitational field determining the metric structure of space-time and defining the concept of inertia is as dynamical entity also represented by ur objects. Thus the existence of space-time in the sense of general relativity appears as a consequence of an underlying quantum theoretical structure interpreted as information.}
\end{figure}

The dynamics and the quantization conditions of the tetrad field have to be considered as a consequence of the dynamics and the quantization of the fundamental spinor fields describing gravity. This is in accordance with the description of gravity within \cite{Kober:2009}. In the present paper these fields are derived from ur objects. Thus the existence of space-time in the sense of general relativity is assumed to be the consequence of a representation of the dynamics of abstract quantum states by ur objects. Since the tetrad field is derived from fundamental spinor fields, the approach is very similar to the theory described in \cite{Kober:2009}, where the gravitational field has its origin in a spin connection of a fundamental spinor field which is itself also expressed by spinor fields from which there is constructed a tetrad field, too. The theory in \cite{Kober:2009} presupposes a background independent setting of Heisenberg's unified quantum field theory of spinors \cite{Heisenberg:1957},\cite{Heisenberg:1967},\cite{Heisenberg:1974du} combined with the idea that a spin connection is more fundamental than the metric field \cite{Dehnen:1986mx},\cite{Ashtekar:1986yd},\cite{Ashtekar:1987gu},\cite{Ashtekar:1989ju}.
The conceptual setting of gravity presented here can be considered as an extension of the one presented in \cite{Kober:2009}.
This is the case in the sense that there the existence of a four dimensional space-time manifold has to be presupposed and just the metric structure is described in the framework of a background independent spinor setting. In the approach of the present paper the existence of a space-time manifold is itself, like the gravitational field, derived from the quantum theory of ur objects. This means that it is not just background independent, it even explains why physical objects can be described by quantum fields on a space-time manifold. Thus there seem to coincide the meaning of diffeomorphism invariance within general relativity which indicates that space-time coordinates just represent coincidences of dynamical entities on the one hand and the nature of quantum theory in its general setting giving rise to the attitude that the existence of a position space as representation space of states is a consequence of its Hilbert space structure and not a prerequisite on the other hand. 
With respect to these conceptual and philosophical interpretations of quantum theory and general relativity this interpretation of the ontological status of space and its relation with time as space-time according to special relativity yields not just a kind of reconciliation of the two theories but it makes explicit their conceptual affinity.
The properties of general relativity presupposing a (3+1)-dimensional space-time endowed with a gravitational field describing geometric qualities of space-time are assumed to be derived from quantum theory in this approach implying that this affinity of the theories has its origin in the fact that they are related to each other. This conceptual hierarchy where the space-time structure of general relativity appears as a consequence of quantum theory interpreted in an abstract sense is represented in figure ($\ref{spacetimequantumstates}$).

\section{Cosmological Considerations}

It has been considered quantum field theory and the incorporation of general relativity in the approach of a quantum theory of ur objects as a representation of abstract quantum theory. Thus the existence of elementary particles and quantum fields appear as consequence of abstract quantum theory. But there still arises the question about the topological structure of the universe and thus the global structure of space-time. So far it has only been obtained a local description of space-time.
Concerning this question there has to be mentioned that within the approach of the quantum theory of ur objects the topological structure of the universe is independent of gravity and thus of the considerations about the incorporation of gravity given above. The topological structure of the universe is directly determined by the ur hypothesis without a reference to a gravitational field \cite{Weizsaecker:1985},\cite{Lyre:1994eg},\cite{Lyre:2003tr}.

Therefore one has to consider again the symmetry group of an ur object, namely the $U(2)=SU(2) \times U(1)$. All physical phenomena have to be interpreted as interactions between ur objects in the presented approach. Therefore the physical state of the universe will not be changed if all ur objects the universe consists of are transformed by the same element of the symmetry group of the ur objects. Thus there is offered the possibility to interpret the space of the symmetry group of the ur objects as space which represents the topology of the universe. This consideration is again in accordance with Einstein's attitude that space-time coordinates just describe coincidences between dynamical entities. Since the group of time translations described by the time evolution operator

\begin{equation}
U(t,t_0)=\exp\left(-iH(t-t_0)\right)
\end{equation}
with the Hamiltonian $H$ as generator represents a $U(1)$ group with the time $t$ as group parameter, the group parameter of the $U(1)$ group appearing above can be interpreted as time coordinate. An arbitrary element of the SU(2) can be represented by 

\begin{equation}
U=\left(\begin{matrix}a+ib&ic+d\\ ic-d&a-ib\end{matrix}\right)\quad,\quad a^2+b^2+c^2+d^2=1.
\end{equation}
Thus the SU(2) corresponds to a $\mathbb{S}^3$ and can be interpreted as the spatial part of the universe.
This means that the topology of the $U(2)$ corresponds to $\mathbb{S}^3$ $\times$ $\mathbb{R}$, if the U(1) group is 
mapped to a real coordinate $\mathbb{R}$ representing time, and according to this it can be identified with the topology 
of the universe.\\
An extended cosmological consideration refers to the tensor space of ur objects which describes states containing many ur objects. This corresponds to the second quantization step in the above classification, described by the postulated commutation relation of the ur objects ($\ref{commutator_ur-object}$) and the introduced creation and annihilation operators for single ur objects ($\ref{creation_annihilation_ur-objects}$). The biggest Lie group which can be represented in an unitary way within the tensor space of ur objects by bilinear expressions of the creation and annihilation operators $a_r$ and $a_r^{\dagger}$ can be described by the following generators 

\begin{eqnarray}
M_{01}&=&\frac{i}{2}(n_1-n_2),\nonumber\\
M_{02}&=&\frac{1}{2}(-\tau_{12}+\tau_{21}),\nonumber\\
M_{12}&=&\frac{i}{2}(\tau_{12}+\tau_{21}),\nonumber\\
M_{34}&=&\frac{i}{2}(n+1),\nonumber\\
\nonumber\\
N_{03}&=&\frac{1}{4}(\lambda_{11}-\lambda_{22}-\lambda_{11}^{\dagger}+\lambda_{22}^{\dagger}),\nonumber\\
N_{13}&=&\frac{i}{4}(\lambda_{11}+\lambda_{22}+\lambda_{11}^{\dagger}+\lambda_{22}^{\dagger}),\nonumber\\
N_{23}&=&-\frac{1}{2}(\lambda_{12}-\lambda_{12}^{\dagger}),\nonumber\\
N_{05}&=&\frac{i}{4}(\lambda_{11}-\lambda_{22}+\lambda_{11}^{\dagger}-\lambda_{22}^{\dagger}),\nonumber\\
N_{15}&=&-\frac{1}{4}(\lambda_{11}+\lambda_{22}-\lambda_{11}^{\dagger}-\lambda_{22}^{\dagger}),\nonumber\\
N_{25}&=&\frac{i}{2}(\lambda_{12}+\lambda_{12}^{\dagger}),\nonumber\\
\end{eqnarray}
where the appearing quantities are defined as follows

\begin{eqnarray}
\lambda_{rs}&=&a_r a_s,\nonumber\\
\lambda_{rs}^{\dagger}&=&a_r^{\dagger} a_s^{\dagger},\nonumber\\
\tau_{rs}&=&a_r^{\dagger} a_s,\nonumber\\
n_r&=&\tau_{rr},\nonumber\\
n&=&\sum_r n_r.
\end{eqnarray}
The M-generators are antisymmetric and the N-generators are symmetric

\begin{equation}
M_{ik}=-M_{ki},\quad N_{ik}=N_{ki}.
\end{equation}
They fulfil the following Lie Algebra

\begin{equation}
[M_{ik},M_{kl}]=M_{il},\quad [N_{ik},N_{kl}]=M_{il},\quad [M_{ik},N_{kl}]=N_{il},\\
\end{equation}
which corresponds to the SO(3,2) group, the Anti de Sitter group leaving the expression

\begin{equation}
s=x_1^2+x_2^2+x_3^2-x_4^2-x_5^2
\end{equation}
invariant. Thus this group describes a four dimensional space being embedded in a five dimensional space. In this sense it can be interpreted as the cosmological space representing the universe if the time coordinate $x_0$ is related to the two coordinates with negative sign in the following way

\begin{equation}
x_4=x_0 \cos{\alpha}\quad,\quad x_5=x_0 \sin{\alpha}.
\end{equation}
Therefore one is led to a realistic cosmological model where the universe consists of ur objects. Both cosmological considerations, the one referring to single ur objects and the one referring to the tensor space of many ur objects, imply a (3+1)-dimensional space-time as global structure describing the universe being in accordance with the derived field equations referring to a (3+1)-dimensional space-time without reference to the global topological structure. This means that within the presented approach the spatial structure of the universe with respect to local relations and with respect to its global shape appears as a direct consequence of the logical structure of abstract quantum theory without any special assumptions belonging to a certain model.

\section{Summary and Discussion}

It has been shown that in the quantum theory of ur objects where the existence of quantum fields in a (3+1)-dimensional space-time is the consequence of a representation of the abstract state space of an arbitrary object interpreted in an information theoretical sense it is also possible to incorporate general relativity in a natural way. This is the case because it is possible to use the translation gauge description of gravity. The tetrad field appearing as gauge potential in such a description is constructed from a couple of spinor fields representing states of the second quantization step of a binary alternative which thus represent states in the tensor space of ur objects. A further quantization of the spinor fields according to the usual setting of quantum field theory leads to a field operator of the tetrad field. 
According to \cite{Kober:2009} the metric structure of space-time arises in a natural way, if one assumes a unified quantum field theory of spinors without presupposing an a priori metric structure. In the present paper the existence of a (3+1)-dimensional manifold as a representation space of relations between quantum states is derived from the representation of quantum states by ur objects whereas in \cite{Kober:2009} it has to be presupposed the existence of a four dimensional manifold and just the metric structure with Lorentz signature is derived by using the mathematics of twistor theory.
The approach presented in this paper takes allowance for a conceptual attitude arising from the principles of quantum theory and general relativity. It is the attitude that the existence of space combined with the time direction in the sense of special relativity has no reality by itself but just as a representation of relations between dynamical entities and that just coincidences of dynamical objects have a real physical meaning.
In general relativity this attitude arises from the property of diffeomorphism invariance or the lack of the existence of absolute objects being related to the fact that the concepts of inertia and acceleration get their meaning by a dynamical entity, the gravitational field namely, which is expressed as tetrad field or metric field. In quantum theory it arises from the property that it is formulated as a theory of abstract Hilbert spaces not presupposing necessarily the existence of a position or a momentum space which can be interpreted as certain representation spaces but are no ontological prerequisites.
In accordance with this quantum fields on a space-time manifold can be interpreted as a kind of representation of underlying abstract quantum states represented by postulated basic entities of quantum theory called ur objects. These ur objects are the simplest objects being thinkable in quantum theory and are therefore assumed to be the fundamental entities of nature. They are no objects in space but constitute space and its symmetry properties and they can be interpreted as basic constituents of information. Since every quantum state is assumed to be composed of a finite number of ur objects corresponding to a finite number of degrees of freedom, there could be offered the possibility to cure the problem of infinities from the outset which arises in usual quantum field theory mandatorily.
The problem of time is not completely treated so far. In the context of quantum theory it seems to be more fundamental than space which does not contradict special relativity since special relativity just asserts that space and time are combined in a new way but not that the difference between them does not exist any more. This is the reason why time is presupposed in the quantum theory of ur objects whereas space and the mathematical relation between space and time is assumed to be a physical reality derived from quantum theory.
It is also important that the existence of special objects like particles is assumed to be already a consequence of abstract
quantum theory without further special postulates belonging to a separated theory of elementary particles.
Thus it seems at least to be possible to derive the existence of elementary particles, space-time, gravity and the topological structure of the universe from quantum theory interpreted as an abstract theory of information assumed to be more fundamental than the existence of matter.\\
\\
$Acknowledgement$:\\ I would like to thank the Messer Stiftung for financial support.

%\newpage

\end{document}